\documentclass[fleqn,usenatbib]{mnras}

\usepackage[T1]{fontenc}
\usepackage{ae,aecompl}

\usepackage{graphicx}	% Including figure files
\usepackage{amsmath}	% Advanced maths commands
\usepackage{amssymb}	% Extra maths symbols

\usepackage{xspace}

\usepackage{subcaption}
\usepackage[percent]{overpic} % Put text over figure

\newcommand{\pow}[1]{\ifmmode{}^{#1}\else ${}^{#1}$\fi}

\newcommand{\Lya}{\ifmmode{\mathrm{Ly}\alpha}\else Ly$\alpha$\xspace\fi}

\newcommand{\cm}{\,\ifmmode{{\mathrm{cm}}}\else cm\fi}

\newcommand{\ergps}{\,{\rm erg}\,{\rm s}\ifmmode{}^{-1}\else ${}^{-1}$\fi}
\newcommand{\Mpch}{\,{\rm Mpc}\,\ifmmode h^{-1}\else $h^{-1}$\fi}
\newcommand{\snru}{\,\ifmmode{\mathrm{Myr}^{-1}}\else Myr${}^{-1}$\fi}
\newcommand{\kms}{\,\ifmmode{\mathrm{km}\,\m $\chi_{100} \equiv (\chi/100)$athrm{s}^{-1}}\else km\,s${}^{-1}$\fi\xspace}

\newcommand{\cl}{\mathrm{cl}}
\newcommand{\mytime}[1]{\ifmmode{{t_{\mathrm{#1}}}}\else $t_{\mathrm{#1}}$\fi}
\newcommand{\tcc}{\mytime{cc}}
\newcommand{\tcool}[1]{\mytime{cool,#1}}
\title[Cold gas in a hot wind]{
  The growth and entrainment of cold gas in a hot wind}

\author[M. Gronke \& S. P. Oh]{
  Max Gronke$^{1}$\thanks{E-mail: maxbg@ucsb.edu}
  and S. Peng Oh$^{1}$
\\
$^{1}$Department of Physics, University of California, Santa Barbara, CA 93106, USA
}

\date{Draft from \today}

\pubyear{2018}

\begin{document}
\label{firstpage}
\pagerange{\pageref{firstpage}--\pageref{lastpage}}
\maketitle

\begin{abstract}
Both absorption and emission line studies show that cold gas around galaxies is commonly outflowing at speeds of several hundred km$\,\textrm{s}^{-1}$.
This observational fact poses a severe challenge to our theoretical models of galaxy evolution since most feedback mechanisms  (e.g., supernovae feedback)
accelerate hot gas, and the timescale it takes to accelerate a blob of cold gas via a hot wind is much larger than the time it takes to destroy the blob.
We revisit this long-standing problem using three-dimensional hydrodynamical simulations with radiative cooling. Our results confirm previous findings, that cooling is often not efficient enough to prevent the destruction of cold gas. However, we also identify regions of parameter space where the cooling efficiency of the mixed, `warm' gas is sufficiently large to contribute new comoving cold gas which can significantly exceed the original cold gas mass. This happens whenever, $t_{\mathrm{cool, mix}}/t_{\mathrm{cc}} < 1$, where $t_{\mathrm{cool,mix}}$ is the cooling time of the mixed warm gas and $t_{\mathrm{cc}}$ is the cloud-crushing time. This criterion is always satisfied for a large enough cloud. Cooling `focuses' stripped material onto the tail where mixing takes place and new cold gas forms. A sufficiently large simulation domain is crucial to capturing this behavior.
\end{abstract}

\begin{keywords}
  galaxies: evolution -- hydrodynamics -- ISM: clouds -- ISM: structure -- galaxy: halo -- galaxy: kinematics and dynamics
  \vspace{-1.5em}
\end{keywords}

%%%%%%%%%%%%%%%%%%%%%%%%%%%%%%%%%%%%%%%%%%%%%%%%%%

%%%%%%%%%%%%%%%%% BODY OF PAPER %%%%%%%%%%%%%%%%%%
\section{Introduction}
\label{sec:intro}
Outflows are ubiquitously detected around galaxies throughout cosmic time. Both absorption and emission line studies of a range of star-forming galaxies (from nearby dwarfs to ULIRGS to high-$z$ LBGs) show offsets of several hundred \kms with respect to the host galaxy \citep[see, e.g.,][and references therein]{Veilleux2005}.
The main source of energy injection leading to these outflows is thought to be supernovae.
However, while the above mentioned observables trace `cold' $\sim 10^4\,$K gas, feedback heats the gas to much higher temperatures. This multiphase structure can also be directly detected in nearby galaxies where the hot gas can be observed through X-ray emission \citep[e.g.,][]{Martin1999}.

These robust observations imply the existence of multiphase outflows in a wide range of astrophysical systems which poses severe challenges to theoretical models, and has been the subject of many studies \citep{Mckee1978,Klein1994,Mellema2002,Pittard2005,Cooper2009,McCourt2015,Scannapieco2015a,Schneider2016}. The reason for this puzzle is that hydrodynamical instabilities (most notably, the Kelvin-Helmholtz and Rayleigh-Taylor instability) mix the phases and ultimately destroy the cold gas. \citet{Klein1994} quantified the destruction timescale of an adiabatic, initially static cloud of size $r_\cl$ with an impinging hot wind of velocity $v_{\mathrm{wind}}$ to be $\tcc \sim \chi^{1/2} r_\cl / v_{\mathrm{wind}}$, of order both the Kelvin-Helmholtz and Rayleigh-Taylor timescales. Here, $\chi$ is the density contrast between the cloud and the wind which typically takes values ranging from $\sim 100$ for galactic to $\sim 1000$ for cluster conditions.
Notably, this `cloud crushing' time $\tcc$ is smaller by a factor of $\chi^{1/2}$ than the `drag time' $t_{\mathrm{drag}}\sim \chi r_{\cl}/v_{\mathrm{wind}}$, which is required for the hot wind to accelerate the cloud via momentum transfer to speed $v_{\mathrm{wind}}$. Thus, reproducing the  cold, dense gas moving at high velocities seen in observations is challenging in hydrodynamical simulations, since cold clouds get destroyed before they can become entrained with the wind. 

Recently, \citet{Zhang2015a} summarized this discrepancy between observations and theory by using hydrodynamical results in the literature combined with an analytic formulation for the outflows which allowed them to derive theoretically feasible outflow speeds of cold gas. They compare their findings to observations, and conclude that hot winds cannot accelerate the cold gas to the observed velocities, or -- as their title suggests -- that the prevailing picture of entrainment is in trouble. Alternatively, non-thermal forces such as radiation pressure \citep{Zhang2018} or cosmic-rays \citep{Wiener2017} could accelerate the cold gas, although this has not yet been shown to be viable. Another possibility is the cold gas forming from an adiabatically cooling wind \citep{Thompson2016,schneider18}.

Here, we revisit the problem of entrainment with an emphasis on certain regions of parameter space. As in some previous studies \citep[e.g.,][and see \S~\ref{sec:prev_work}]{Mellema2002} we include radiative cooling, which prolongs the lifetime of the cold gas. We focus on the limit of fast cooling rates, i.e., when the cooling time of the cloud $\tcool{cl}\ll\tcc$. Within this regime, we identify a region in parameter space which not only allows the cold gas to survive until entrainment, but also enables new, cold gas to condense out of the hot wind.

\begin{figure*}
  \centering
  \vspace{-.2cm}
  \includegraphics[width=.94\linewidth]{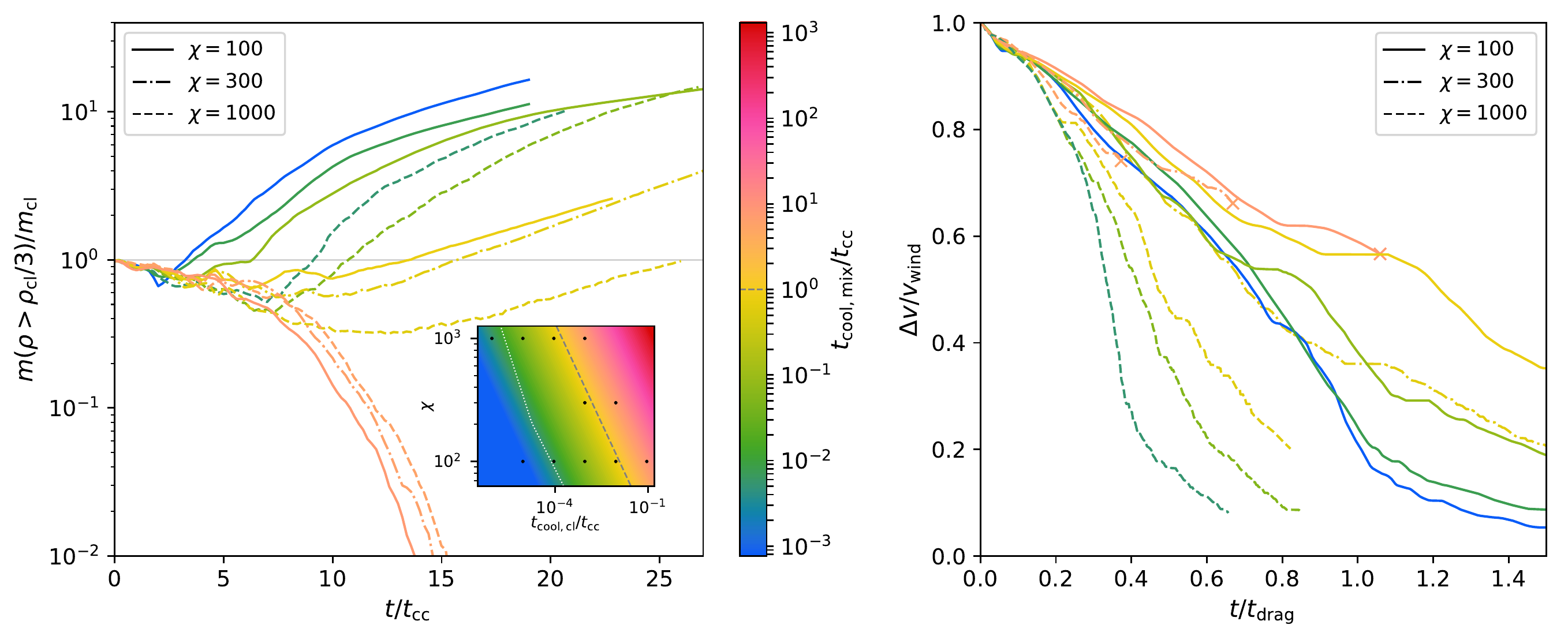}
  \vspace{-.35cm}
  \caption{\textit{Left panel:} Mass evolution for three different cloud densities (denoted by the linestyle). We show the mass contained in dense gas ($\rho > \rho_{\cl}/3$). The color coding of the curves indicates $\tcool{mix}/\tcc$. In the inset, we show how the overdensity $\chi$ and the cloud cooling time $\tcool{cl}/\tcc$ relate to this fraction, and mark our simulations with black dots. The grey dashed line in the figure's inset denotes the threshold $\tcool{mix}/\tcc = 1$ above and below which the cloud loses and gains mass, respectively. The dotted white line marks the threshold $t_{\rm cool, wind}\sim t_{\rm drag}$, i.e., the conservative limit left of which the wind itself will start to cool without an additional heating source.
    \textit{Right panel:} Evolution of the velocity difference between hot and cold phase. The color coding and linestyles match the left panel but the time is rescaled to $t_{\mathrm{drag}}=\tcc \chi^{1/2}$ instead. We display the evolution until the end of the simulation, or when the dense gas mass $m(\rho > \rho_\cl / 3)$ falls below $0.1 m_\cl$, in which case we mark the point with a cross.}
  \label{fig:mass_evolution}
  \vspace{-.6em}
\end{figure*}

\vspace{-1.5em}
\section{Analytical considerations}
\label{sec:analytics}
Intuitively, one might expect that in the rapid cooling regime, it is more difficult to destroy the cold gas cloud as has been pointed out \citep{Klein1994,Mellema2002}. However, recent detailed simulations conclude that although cooling extends the lifetime of the cloud, the cold gas mass still decreases and mixes away before the cloud becomes entrained \citep{Cooper2009,Scannapieco2015a,Schneider2016}.

The conversion from cold to hot gas occurs predominantly via mixing due to the Kelvin-Helmholtz instability. \citet{Begelman1990} describe the behavior of such a turbulent mixing layer analytically (see also Ji et al 2018, in preparation).
They estimate the mass flux from the wind and the cloud to the mixing layer to be $\dot m_{\mathrm{wind}}\sim \eta_{\mathrm{wind}} v  \rho_{\mathrm{wind}}$ and $\dot m_{\cl} \sim \eta_\cl \rho_\cl l / t_{\mathrm{KH}}(l) \sim \eta_{\cl} (\rho_\cl \rho_{\mathrm{wind}})^{1/2}v$, respectively. Here, $v$ and $l$ are the characteristic speed and length scale of the mixing process, and $\eta_{i}$ denotes the deposition efficiency. Note that since $t_{\mathrm{KH}}\sim \chi^{1/2} l/v$, the mass flux from the cold medium into the mixing layer is independent of its extent.

With these mass fluxes into the mixing layer, the typical mixing temperature is then \citep[following][]{Begelman1990}
\begin{equation}
  \label{eq:Tmix}
  T_{\mathrm{mix}}= \frac{\dot m_\cl T_\cl + \dot m_{\mathrm{wind}} T_{\mathrm{wind}}}{\dot m_\cl + \dot m_{\mathrm{wind}}} \sim \sqrt{T_{\mathrm{wind}}T_\cl}\;,
\end{equation}
where the last equality is true if $\eta_{\cl}\sim\eta_{\mathrm{wind}}$. Under the assumption of pressure equilibrium, the corresponding mixing density is $\rho_{\mathrm{mix}}\sim \chi^{1/2}\rho_{\mathrm{wind}}$. This leads to a cooling time of the mixed gas $\tcool{mix} \sim \chi \frac{\Lambda(T_\cl)}{\Lambda(T_{\mathrm{mix}})} \tcool{cl}$ where $\Lambda(T_\cl) / \Lambda(T_{\mathrm{mix}})\sim 0.1-1$ for $4.2 \lesssim \log_{10}T_\cl \lesssim 4.6$.
This implies that the fate of the cloud is controlled by the cooling time of the mixed gas.
If $\tcool{mix} < \tcc$, we expect the mixing layer to cool sufficiently fast to replenish the cold gas reservoir, and thus, extend the lifetime of the cloud. 
In addition, the newly cooled gas (part of which originated in the hot wind)  already carries high momentum, leading to faster entrainment.

The requirement $\tcool{mix}/\tcc < \alpha$ corresponds to a cloud of size
\begin{equation}
R > \frac{v_{\rm wind}\tcool{mix}}{\chi^{1/2}} \alpha^{-1} \approx 2 \, {\rm pc} \ \frac{T_{\rm cl,4}^{5/2} \mathcal{M}_{\mathrm{wind}}}{P_{3} \Lambda_{\rm mix,-21.4}} \frac{\chi}{100}\alpha^{-1}
\end{equation}
where $T_{\rm cl,4} \equiv (T_{\rm cl}/10^{4} \, {\rm K})$, $P_{3} \equiv nT/(10^{3} \, {\rm cm^{-3} \, K})$, $\Lambda_{\rm mix,-21.4} \equiv \Lambda(T_{\rm mix})/(10^{-21.4} \, {\rm erg \, cm^{3} \, s^{-1}})$, $\mathcal{M}_{\mathrm{wind}}$ is the Mach number of the wind, and we write $v_{\mathrm{wind}} = c_{\mathrm{s,wind}} \mathcal{M}_{\rm wind} \sim c_{\mathrm{s,cl}}\mathcal{M}_{\rm wind}\chi^{1/2}$. This implies that for typical galactic conditions clouds larger than parsec size usually fulfill the requirement.

The scale $R$ can be compared to the characteristic length scale of cooling induced fragmentation $l_{\mathrm{M}}\sim c_{\mathrm{s,cl}}t_{\mathrm{cool,cl}}$ \citep{McCourt2016} to give
\begin{equation}
\frac{R}{l_{\mathrm{M}}} \approx 10 \ \mathcal{M}_{\rm wind} \frac{\chi}{100} \left(
\frac{\Lambda(T_{\rm cl})/\Lambda(T_{\rm mix})}{0.1} \right) \alpha^{-1}.
\end{equation}
Thus, a cloud should be significantly larger than the characteristic fragmentation mass to exhibit this behavior. Note that even a swarm of droplets of size $\sim l_{\mathrm{M}}$ subject to a wind can coagulate to form larger clumps \citep{McCourt2016}.

\vspace{-1.5em}
\section{Numerical Method}
\label{sec:method}
We use \texttt{Athena} $4.0$ \citep{Stone2008ApJS..178..137S} to solve the three-dimensional hydrodynamical equations on a uniform, Cartesian grid. Our simulation setup corresponds to a classical `cloud crushing' scenario, and is similar to others. Specifically, we place a spherical, dense, cold clump with radius $r_{\cl}$ in our simulation domain and impose a steady, hot wind with Mach number $\mathcal{M}_{\mathrm{wind}}=1.5$ outside the clump. Such a Mach number corresponds to the transonic outflows seen in galactic winds; we also later consider higher Mach number flows. Our boundaries are all outflowing, except in the upstream direction where we impose a constant inflow of hot gas with density $\rho_{\mathrm{wind}}$.
We use density ratios between the cloud and the wind of $\chi\equiv\rho_{\cl}/\rho_{\mathrm{wind}}=\{100,\,300,\,1000\}$ to cover the parameters relevant for a range of astrophysical systems, and initialize the setup in pressure equilibrium with the initial cloud's temperature to be $T_\cl\approx 4\times 10^4\,$K.\footnote{Gas in photoionization equilibrium with the EUV background can be a factor of two cooler; we later explore the effect of different temperature floors, and $T_{\cl}$.} As in \citet{McCourt2015} we use a cloud-tracking scheme which changes the reference frame continuously to follow the cold gas. In spite of this, we choose fairly large boxes with sizes ranging from $60 r_\cl$ to $400 r_\cl$ along the wind direction, and $12 r_\cl$ to $18 r_\cl$ orthogonal to it for reasons which will become clear below. The main goal of this work is to approximately quantify the evolution of the global cold gas mass for various parameters.
We therefore focused on a suite of low resolution simulations with $r_{\cl}/d_{\mathrm{cell}}=8$ but we also performed convergence tests with simulations up to $r_{\cl}/d_{\mathrm{cell}}=64$ which has been adopted in the most recent $3$D cloud-crushing studies \citep[e.g.,][]{Schneider2016}.

We implement cooling using the \citet{Townsend2009} algorithm with a $7$ piece power-law fit to the \citet{Sutherland1993} cooling curve. Furthermore, we impose a temperature floor at $T_{\mathrm{floor}}=T_\cl$. This measure mimics a heating source, and we tested its impact by lowering $T_{\mathrm{floor}}$, changing the cooling curve to a $67$ piece power law fit of the collisional ionization equilibrium cooling curve used in \citet{Wiersma2009}\footnote{This cooling curve was used by \citet{Scannapieco2015a}.}, and introducing a heating source. We present these test results alongside with convergence studies, and higher Mach number runs in \S~\ref{sec:caveats}.

For our parameter study, we varied both $\chi$ and the cloud radius by changing $\tcool{cl}/\tcc$ from $\sim 10^{-6}$ to $10^{-1}$ where $\tcool{cl}$ is the initial cooling time of the cloud.

\begin{figure*}
  \centering
  \input{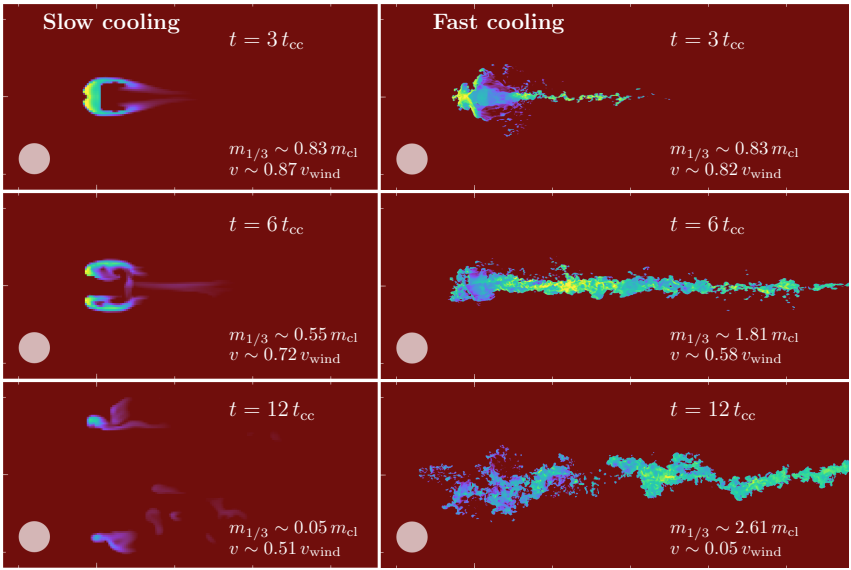}
  \parbox[c]{.23\linewidth}{%
    \caption{
      Density contours of $\chi=100$ simulations with $\tcool{mix}/\tcc\sim 7.7$ (left column), and $\tcool{mix}/\tcc\sim 0.077$ (right column) at $t/\tcc = \{3,\,6,\,12\}$ (from top to bottom row). The simulation displayed on the left (right) was performed with a resolution of $r_\cl / d_{\mathrm{cell}}=8$ ($r_\cl / d_{\mathrm{cell}}=64$). The grey disk in each panel illustrates the original size of the cloud. For visualization purposes we display only part of the box in the slow cooling simulation (left column) but checked that no gas is forming outside this boundary.
      \label{fig:multi2d}
    }}
  
\vspace{-1\baselineskip}
  
\end{figure*}

\begin{figure}
  \centering
  \vspace{-.1cm}
  \includegraphics[width=\linewidth]{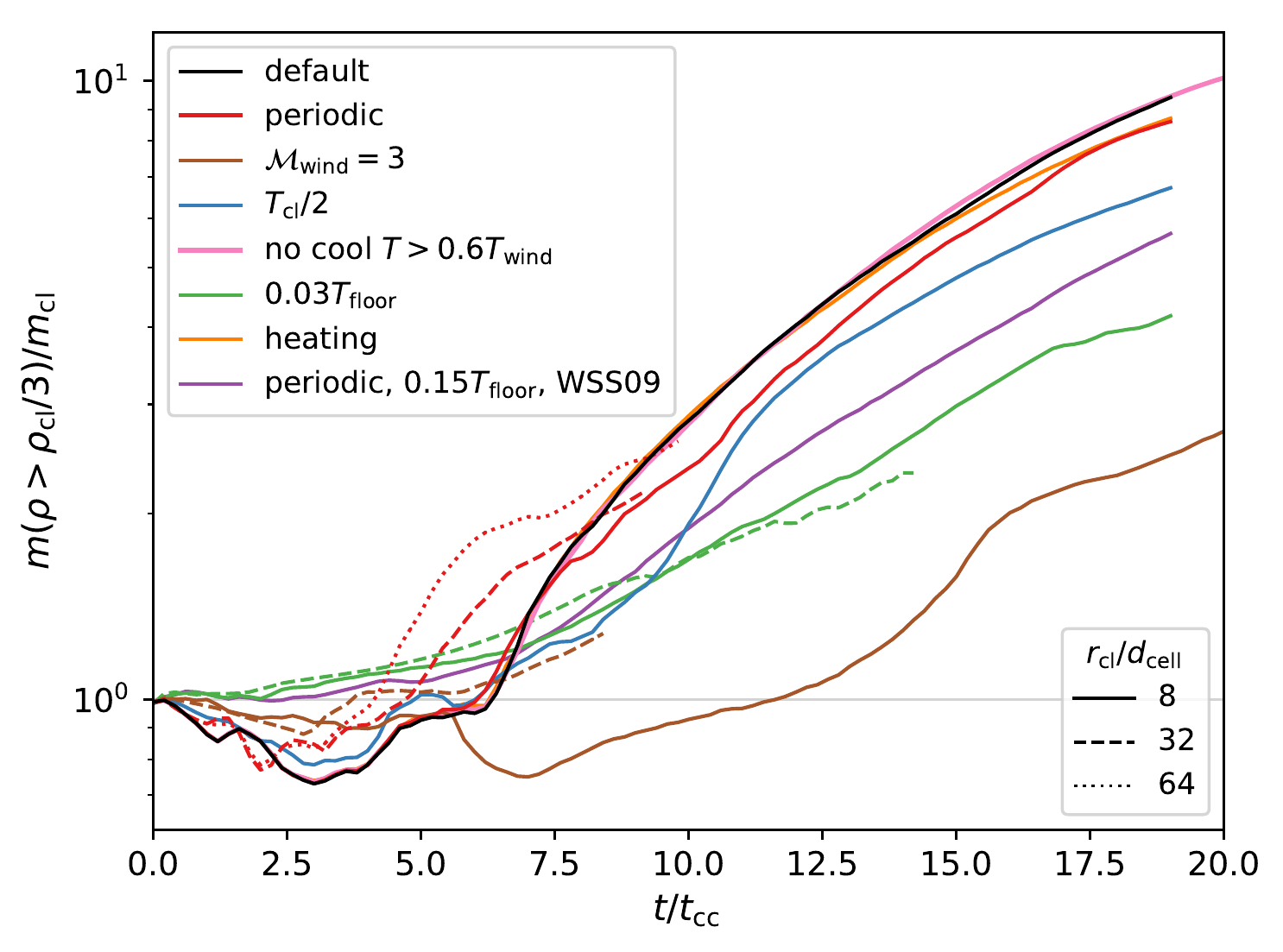}
  \vspace{-.7cm}
  \caption{Mass evolution for $\tcool{cl} / \tcc \sim 10^{-3}$ with $\chi=100$ and several other changed input parameters such as increased resolution (different linestyles), altered boundary conditions (`periodic', \textit{red}) or higher Mach number (\textit{brown}).
    The \textit{green} and \textit{blue} lines show the impact of lowered cooling floor (to $0.03 T_\cl$) and reduced $T_\cl$, respectively. The runs represented with the \textit{purple} and \textit{orange lines} use an alternative cooling function, and include a simplistic form of heating, respectively. In \textit{pink} we show the effect of a switched off cooling for $T>0.6 T_{\mathrm{wind}}$.}
  \vspace{-1.5em}
  \label{fig:d100_check}
\end{figure}

\vspace{-1.5em}
\enlargethispage{0.2\baselineskip}
\section{Results}
\label{sec:results}
The left panel of Fig. \ref{fig:mass_evolution} shows the dense ($\rho > \rho_\cl/3$) mass evolution of simulations with three different overdensities $\chi$ (marked with different linestyles; all normalized to the initial cloud mass $m_\cl$), and several cloud sizes. The color of each curve denotes the ratio $\tcool{mix} / \tcc$, and the inset shows how our simulation parameters relate to this ratio. Notably, simulations with $\tcool{mix} / \tcc \gtrsim 1$ lose mass (with $\lesssim 1\%$ of the initial cloud mass left by $t\sim 15\tcc$ independent of $\chi$), while the runs with $\tcool{mix} / \tcc \lesssim 1$ not only retain the initial cloud mass but exceed this by up to an order of magnitude. Initially all cold gas masses drop for $\sim 0.2 t_{\mathrm{drag}}$ before stabilizing, and then entering a phase of mass growth. The rate of the mass growth depends on the cooling rate. However, it seems to reach a limit for $\tcool{mix} / \tcc \rightarrow 0$, perhaps because the rate is saturated by the efficiency of the mixing (and, thus, $\tcc$).
In upcoming work, we will study how mass growth eventually shuts off.

The background wind cools if the cooling rates are very large, and no additional heating source is provided. A conservative limit for when this occurs is $\tcool{wind} \lesssim t_{\mathrm{drag}}$ which is marked by the white dashed line in the inset of Fig.~\ref{fig:mass_evolution}.
In order to analyze the limit $\tcool{cl}\rightarrow 0$ we performed simulations where the wind cools (marked by the two points to the left of the white dashed line in the inset of Fig.~\ref{fig:mass_evolution}). We therefore set the cooling efficiency to zero for $T > 0.6 T_{\mathrm{wind}}$, and to allow for a fair comparison, we do so for all the runs presented in Fig.~\ref{fig:mass_evolution}. Note that all other runs presented subsequently do not have this switch on; Fig.~\ref{fig:d100_check} also shows explicitly that there is no impact on our default run. The inset of Fig.~\ref{fig:mass_evolution} also shows the location of our simulations on the $(\tcool{cl}/\tcc,\,\chi)$-plane (black circles), and (as a dashed line) the $\tcool{mix} / \tcc = 1$ threshold. The simulations close to this threshold -- but still with $\tcool{mix} / \tcc < 1$ gain mass very slowly, with the $\chi=1000$ simulation reaching $m(\rho > \rho/3) \sim m_\cl$ only after $\sim 25\tcc$.

The right panel of Fig.~\ref{fig:mass_evolution} shows the corresponding velocity evolution of the cold gas. Specifically, we plot the velocity difference between the reference frame set by the dye weighted velocity and the hot wind, versus the time normalized by the expected acceleration time, $t_{\mathrm{drag}}=\chi^{1/2}\tcc$.
We see that entrainment indeed happens on the characteristic timescale of $\sim t_{\mathrm{drag}}$, although faster cooling leads to quicker entrainment. Indeed, for the higher overdensities, the entrainment time appears significantly shorter than $t_{\mathrm{drag}}$. The rapid reduction in shear arises from the cooling of already comoving hot gas, rather than rapid acceleration of pre-existing cold gas. For instance, for $\chi=1000$ the rapid drop in $\Delta v$ over $t/t_{\mathrm{drag}}\sim 0.2-0.5$ ($t/\tcc \sim 6-15$) corresponds to a period when the cold gas mass grows to $\sim 10$ times its initial value. In this limit, the cold gas velocity asymptotes to $v\rightarrow \frac{m_{\mathrm{cold}}}{m_{\mathrm{cold}} + m_{\cl}} v_{\mathrm{wind}}$, as expected for an inelastic momentum conserving collision, and thus $\Delta v/v_{\mathrm{wind}} \rightarrow \frac{m_\cl}{m_{\mathrm{cold}}+m_{\cl}}$.

In order to understand the physical process of mass growth and entrainment, we show in Fig.~\ref{fig:multi2d} the evolution of the density slices from snapshots at $t=\{3,\,6,\,12\}\tcc$ (from top to bottom row) of
simulations with two different cloud radii ($\tcool{mix} / \tcc \sim 8$ in the left and $\tcool{mix} / \tcc \sim 0.08$ in the right column). The two simulations displayed have a resolution of $r_\cl / d_{\mathrm{cell}}=8$ and $r_\cl / d_{\mathrm{cell}}=64$ shown in the left and right column of Fig.~\ref{fig:multi2d}, respectively. The  slow cooling case is already well-studied in the literature, so we only run a low-resolution simulation.With increasing resolution, the slowly cooling cloud mixes and is destroyed even faster \citep[e.g.,][]{Schneider2016}.

The first row of Fig.~\ref{fig:multi2d} shows that in both cases a reverse shock is driven into the cloud. However, in the fast cooling regime, the mixed gas in the tail of the cloud cools and grows denser. In contrast, in the slow cooling case, the tail mixes quickly and the density falls below $\sim \rho_\cl / 10$. The detailed stages of this process have already been described in previous work \citep[e.g.,][]{Klein1994,Pittard2005,Cooper2009}. Instead, we focus on the dynamics of the fast-cooling case (right column of Fig.~\ref{fig:multi2d}) where we can see the tail only slightly visible at $t\sim 3\tcc$ gaining mass until its thickness reaches approximately the initial cloud extent and its density is $\sim \rho_\cl$ (at $t\sim 6\tcc$).
As warm gas in the tail cools, it creates thermal pressure gradients which further draws hot wind gas into the tail, mixing, and cooling the gas. This creates a cooling-induced focusing effect whereby the low-entropy cold gas stripped off the sides of the cloud is redirected toward the cooling tail, ensuring its survival. This is in contrast to the adiabatic cloud-crushing case where stripped material travels away from the cloud axis and is mixed away.

Importantly, this `reformation point', where the tail forms and thickens, is a few cloud radii downstream of the cloud. This has implications for the required size of the simulation box (in spite of the cloud tracking scheme used) which we discuss in \S~\ref{sec:prev_work}.
If the box is too small, the cooling tail is not well captured and the cloud appears to disintegrate.

\vspace{-1.5em}
\section{Discussion}
\label{sec:discussion}

\subsection{Comparison to previous work}
\label{sec:prev_work}
Cloud crushing has been extensively studied previously. However, most work focused on adiabatic simulations \citep[e.g.][]{Klein1994,Pittard2005}. Other authors include cooling \citep{Cooper2009,Scannapieco2015a,McCourt2015,Schneider2016} but still find destruction of the cold gas mass prior to entrainment.

We attribute this apparent discrepancy with our findings to two reasons: either the cooling was not efficient enough (i.e., the simulated clouds were too small), or the chosen boxsize was not sufficient. Regarding the first point, we computed the cooling times for the parameters used in the literature and found that \citet{McCourt2015} and \citet{Schneider2016} used $\tcool{mix} / \tcc \gtrsim 1$ but the runs of \citet{Cooper2009} and \citet{Scannapieco2015a} fulfilled $\tcool{mix} / \tcc \ll 1$.

However, as described in \S~\ref{sec:results} it is crucial to include the tail -- where the cold gas reforms -- in the simulation domain. Even for very efficient cooling, a lower (upper) estimate on the required boxsize is given by the distance the mixed material travels before the cloud gets destroyed (becomes comoving), i.e., $\tcc v_{\mathrm{wind}}$ ($t_{\mathrm{drag}}v_{\mathrm{wind}}$) which in units of $r_\cl$ is $\chi^{1/2}$ ($\chi$). Since the wind pushes the cloud only a distance of $v_{\mathrm{wind}}/t_{\mathrm{drag}} \tcc^2 \sim r_\cl$ before destruction, the lower limit is required even when utilizing a moving reference frame as we do. In practice, we find tail lengths of $\sim 40 r_\cl$ and $\sim 250 r_\cl$ for our runs with $\chi=100$ and $\chi=1000$, respectively, which are closer to our upper limit. 
These boxsize requirements stem from our runs for $\tcool{cl} / \tcc \ll 1$ and are larger if the cooling is less efficient.
\citet{Scannapieco2015a} used a cloud tracking scheme but their simulation domain from the cloud center to the downstream border is only $8 r_\cl$ -- too low to capture these effects (their runs have $\chi \ge 300$).
\citet{Cooper2009}, on the other hand, did not use a cloud tracking scheme but a long box of length $\sim 30 r_\cl$. However, as they do not employ a moving reference frame this is also too small. In fact, for our $\tcool{mix} / \tcc \sim 10^{-3}$ and $\chi=1000$ run which comes closest to their setup, we find a tail length of $\sim 250 r_\cl$ and we shifted our reference frame by $\sim 45 r_\cl$ at $t\sim 10\tcc$, i.e., when the mass growth starts.

Some previous work has studied cloud-crushing in conjecture with cooling (mainly in the context of gas accretion onto the galaxy), and found mass increase in the wake of the cloud -- similar to our results. \citet{Marinacci2010} found a few $\sim \%$ increase in their cold gas mass over the evolution of their 2D simulations. \citeauthor{Armillotta2016} \citeyearpar{Armillotta2016,Armillotta2017} found in 2D simulations that cloud mass loss slows as cloud size increases, and also show a 3D run with a  $\sim 20\%$ mass increase. Here, we formulate a quantitative criteria for cloud growth which matches a parameter sweep in 3D, and shown that even more substantial growth (by an order of magnitude) is possible.

\vspace{-1.em}
\subsection{Caveats and future work}
\label{sec:caveats}
\textit{Resolution.} Our fiducial resolution of $r_\cl / d_{\mathrm{cell}}=8$ is too low to capture the detailed physical evolution of the problem. In fact, \citet{Klein1994} and \citet{Nakamura2006} suggest $r_\cl / d_{\mathrm{cell}}\gtrsim 100$ are necessary (in $2$D), but recent work finds that even for $r_\cl / d_{\mathrm{cell}} > 1000$ the solution is not converged \citep{Yirak2010} in terms of, e.g., the shock structure. Indeed, \citet{McCourt2016} have previously argued that resolving the lengthscale $l_{\mathrm{M}}\sim c_{\mathrm{s}}t_{\mathrm{cool}}$ is required to resolve the detailed morphology and dynamics of the gas. We barely resolve $l_{\mathrm{M}}$ in some of our simulations for the mixed gas, and not at all in the cold gas.
In Fig.~\ref{fig:d100_check} we compare the global mass evolution of our `default' $\chi=100$ run with $\tcool{cl} / \tcc \sim 10^{-3}$ to resolutions up to $r_\cl / d_{\mathrm{cell}}=64$ which is commonly used for $3$D cloud crushing simulations (e.g., \citealp{Schneider2016}; also \citealp{Scannapieco2015a} used an adaptive technique with a maximum refinement corresponding to this resolution).
The mass evolution differs in detail, but all three show mass doubling by $t\sim 8\tcc$. Note that mass growth is faster at higher resolution suggesting that it is driven by mixing and that we conservatively underestimate it.
Of course, we cannot rule out that increasing the resolution even further will not change the evolution drastically.
However, \citet{McCourt2016} show in $2$D that a collection of small cloudlets can survive and entrain in a wind (see their figure~6).
Addressing this issue in $3$D requires computational resources beyond the scope of this Letter.

\textit{Boundary conditions; cooling.} Fig.~\ref{fig:d100_check} also shows the impact of periodic (versus the default choice of outflowing) boundary conditions orthogonal to the flow -- which lead to a slightly lower mass after $t\gtrsim 7\tcc$ due to the re-entering shock fronts perturbing the tail. It also shows the impact of the cooling function. While the reduction of the initial cloud temperature does not seem to have a strong effect, the reduction of the temperature floor changes the evolution significantly. Firstly, the initial mass loss disappears because of the initial cooling (and contraction) of the cloud. Secondly, the subsequent mass growth is slower which is due to the now higher overdensity of the cloud, which increases the mixing timescales. Indeed, the mass evolution mimics that of a cloud with higher initial overdensity $\chi$.
Similarly, if we change the cooling curve to a fit to the $Z\sim Z_\odot$ collisional ionization cooling function provided by \citet{Wiersma2009} (with a lower temperature floor), we find the overall mass evolution changed, but the effect of mass growth is still present. Also, if we include a simplistic heating model in our simulation by distributing the radiated energy at each timestep uniformly over the simulation domain, or if we switch off cooling for $T > 0.6 T_{\mathrm{wind}}$ the evolution does not change.

\textit{Wind speed.} Even higher wind Mach numbers can occur in the driving region of classic wind models \citep{chevalier85}. Fig.~\ref{fig:d100_check} shows the mass evolution of $\mathcal{M}_{\mathrm{wind}}=3$ runs. Here, a bow shock compresses the cloud and its tail, leading to higher overdensities and hence slower mixing and growth, but the qualitative trend is the same.

\textit{Additional physics} which we ignored in our work include: less idealized cloud and wind properties, anisotropic thermal conduction (see \citet{Bruggen2016, Armillotta2016} for the isotropic case), magnetic fields, a more realistic cooling curve which includes heating. We plan to investigate such issues in future work. 

\vspace{-1.em}
\section*{Acknowledgments}
\vspace{-0.3em}
We thank S. Ji, N. Kulkarni, M. McCourt, and A. Praturu for helpful conversations, and sharing their code with us. We also thank E. Schneider, S. Shen, R. Wissing for interesting discussions. We thank the law offices of May Oh \& Wee, where part of this work was carried out, for hospitality.  
We acknowledge support from NASA grant NNX17AK58G, XSEDE grant TG-AST140047, and the Texas Advanced Computing Center (TACC) of the University of Texas at Austin. This research made use of  \texttt{yt} \citep{2011ApJS..192....9T}.
\vspace{-2em}



\bibliographystyle{mnras}
\bibliography{references}

%%%%%%%%%%%%%%%%%%%%%%%%%%%%%%%%%%%%%%%%%%%%%%%%%%

%%%%%%%%%%%%%%%%% APPENDICES %%%%%%%%%%%%%%%%%%%%%

%\appendix


% Don't change these lines
\bsp	% typesetting comment
\label{lastpage}

\end{document}